\documentclass[preprint,12pt,3p]{elsarticle}
\usepackage{amssymb}
\usepackage{booktabs}
\usepackage{dcolumn}
\usepackage{multirow}
\usepackage{adjustbox}
\usepackage{subfig}
\usepackage{graphicx}
\usepackage{natbib} 
\usepackage{booktabs}
\usepackage{lscape}
\usepackage{amsmath}
\usepackage{float}
\usepackage[obeyspaces,hyphens]{url}
\usepackage{xcolor}
\usepackage{setspace}
\usepackage{indentfirst}
\usepackage{wasysym}
\usepackage{booktabs}
\usepackage{siunitx}
\usepackage{amssymb}
\usepackage{float}
\usepackage[countmax]{subfloat}
\biboptions{sort&compress}
\newcommand{\be}{\begin{equation}}
\newcommand{\ee}{\end{equation}}
\newcommand{\beq}{\begin{eqnarray}}
\newcommand{\eeq}{\end{eqnarray}}
\newcommand{\lb}[1]{\label{#1}}

\newcommand{\ssty}{\scriptscriptstyle}
\newcommand{\tsty}{\textstyle}

\newcommand{\dl}{d_{\ssty L}}

\newcommand{\dg}{d_{\ssty G}}

\newcommand{\gaml}{\gamma_{\ssty L}}

\newcommand{\dobs}{d_{{\tsty {\ssty \rm obs}}}}
\newcommand{\Nobs}{N_{{\tsty {\ssty \rm obs}}}}
\newcommand{\Vobs}{V_{{\tsty {\ssty \rm obs}}}}
\newcommand{\gobs}{\gamma_{{\tsty {\ssty \rm obs}}}}

\newcommand{\Dl}{D_{\ssty L}}
\newcommand{\Dg}{D_{\ssty G}}

\makeatletter
\newcommand{\oz}{\mathbin{\mathpalette\make@circled{z}}}
\newcommand{\odl}{\mathbin{\mathpalette\make@circled{\dl}}}
\newcommand{\ogams}{\mathbin{\mathpalette\make@circled{\gaml^\ast}}}
\newcommand{\make@circled}[2]{%
  \ooalign{$\m@th#1\smallbigcirc{#1}$\cr\hidewidth$\m@th#1#2$\hidewidth\cr}%
}
\newcommand{\smallbigcirc}[1]{%
  \vcenter{\hbox{\scalebox{1.6}{$\m@th#1\bigcirc$}}}%
}
\makeatother
\usepackage{array}

\journal{}

\begin{document}

\begin{frontmatter}

\title{Fractal dimension of the cosmic web with different galaxy types}

\author[1]{Ana Elisa Lima\fnref{fn1}}
\address[1]{Chemistry Institute, Universidade Estadual de Campinas,
	   Campinas, Brazil}
\ead{a204409@dac.unicamp.br}

\author[2]{Julianne C.\ Soares\fnref{fn2}}
\address[2]{Geosciences Institute, Universidade Federal do Rio de
           Janeiro, Rio de Janeiro, Brazil}
\ead{juliannesoares@ufrj.br}

\author[3]{Ana Carolina S.\ Tavares\fnref{fn3}}
\address[3]{Faculty of Animal Science and Food Engineering, Universidade
de S\~{a}o Paulo, S\~{a}o Paulo, Brazil}
\ead{tavares.carolina@usp.br}

\author[4]{Mariana V.\ Taveira\fnref{fn4}}
\address[4]{Division of Physics, Mathematics and Astronomy,
California~Institute~of~Technology,~USA}
\ead{mtaveira@caltech.edu}

\author[5]{Sharon Teles\fnref{fn5}}
\address[5]{Physics Institute, Universidade Federal do Rio de Janeiro,
	   Rio de Janeiro, Brazil}
\ead{steles.ts@gmail.com}

\author[6,7]{Amanda R. Lopes\fnref{fn6}}
\address[6]{Instituto de Astrof\'{\i}sica de La Plata, CONICET-UNLP,
La Plata, Argentina}
\address[7]{Institute of Astronomy, Geophysics, and Atmospheric Sciences,
Universidade de S\~{a}o Paulo, S\~{a}o Paulo, Brazil} 
\ead{amandalopes1920@gmail.com}

\author[5]{Marcelo B.\ Ribeiro\fnref{fn7}\corref{cor3}%
}
\ead{mbr@if.ufrj.br}

\cortext[cor3]{\it Corresponding author}
\fntext[fn1]{Orcid 0009-0007-9041-9014}
\fntext[fn2]{Orcid 0009-0005-6305-4860}
\fntext[fn3]{Orcid 0009-0002-8403-6299}
\fntext[fn4]{Orcid 0009-0002-8012-390X}
\fntext[fn5]{Orcid 0000-0003-4497-9161}
\fntext[fn6]{Orcid 0000-0002-6164-5051}
\fntext[fn7]{Orcid 0000-0002-6919-2624}

\begin{abstract}
The fractal dimension $D$ is used to map the large-scale galaxy distribution
in the Universe by color types: blue, green and red. Using a $NUVrK$-complete
COSMOS2020 subsample of 618,952 galaxies observed up to $z=4$, number densities
were derived and plotted against two cosmological distance measures, the
luminosity and comoving (galaxy area) distances, in order to estimate $D$
for each galaxy color type in two redshift intervals: $z\gtrless1$. We found
a general gradient $D_{\mathrm{blue}}> D_{\mathrm{red}}>D_{\mathrm{green}}$ with
$D=1.40-2.03$ for $z<1$. For $1<z\leq4$, the gradient changes to
$D_{\mathrm{blue}}>D_{\mathrm{green}}>D_{\mathrm{red}}$, and the fractal
dimension values are lower, $D=0.03-0.44$. These results suggest that the
fractal dimension is a sensitive diagnostic for how galaxy populations trace
the evolving cosmic web, and confirm the fractal dimension as a useful tool
for observational mapping of large-scale structure by galaxy color.
\end{abstract}


\end{frontmatter}


\section{Introduction}\lb{intro}

Galaxies are distributed in a highly clustered and void-dominated fashion,
an irregular pattern that makes fractal geometry attractive for describing
large scale topology and, possibly, galaxy evolution. Galaxies have various
shapes, sizes, and colors, then to try to understand how their morphologies
are related to their physical features requires classification in some way
\cite{REF1}. However, the breadth of galaxy properties makes galaxy 
classification challenging.

Advances in observational technology in recent decades have supplied
researchers with galaxy redshift surveys reaching distances previously
inaccessible. Large surveys have mapped the Universe with increasing
details, revealing a structurally rich cosmos and providing critical data
for testing cosmological models. One of these tests is if the galaxy
distribution inhomogeneous pattern would become homogeneous at larger
scales, a subject where fractal geometry can be useful as it provides
ways of characterizing very irregular structures \cite{mandelbrot83}.
This is so because the single fractal dimension $D$ measures the degree
of inhomogeneity since the irregularly distributed galaxy system is
viewed as self-similar, and, therefore, forming a fractal system in
which $D$ quantitatively depicts galactic clustering sparsity or,
complementarily, the dominance of voids in the large-scale structure
of the Universe \cite{pietronero87,coleman92,ribeiro98}, even
connecting with galaxy mergers \cite{bruno2025}. Fractals have so far
been mainly used in cosmology to discuss this problem, with arguments pro
\cite{sylos98,gabrielli2005,gabriela, Teles2021,Teles2022,Teles2023} and
against \cite{scrimgeour2012,goncalves2021} an unlimited galaxy fractal
system up to our present observations, and possibly beyond, although a
theoretical prediction that the standard cosmological model does allow for
unlimited \textit{observational} inhomogeneity was advanced some decades
ago \cite{ribeiro92b,ribeiro95,ribeiro2001a,ribeiro2001b,ribeiro2005,
juracy2008}.

The aim of this work is to use fractal geometry for galaxy clustering
classification. So, here we shall not focus on the problem of if
observations indicate an unlimited, irregularly distributed, fractal
galaxy system. The starting point of this study was a by-product of
Ref.\ \cite{Teles2022} showing that galaxies with different colors
exhibit different single fractal dimensions at different redshift
ranges once their distribution is constructed using cumulative number
counts vis-\`{a}-vis various relativistic cosmological distances. This
by-product revealed that in the COSMOS2015 survey red galaxies have
single fractal dimensions $D$ bigger than blue galaxies for $z<1$,
whereas the situation is the opposite for $z>1$. For the SPLASH survey
blue galaxies have $D$ bigger than red galaxies \cite{Teles2022}. So,
the aim here is to further analyze this feature with a much bigger
and up-to-date galaxy survey in order to confirm the fractal dimension
as a sensitive diagnostic for how galaxy population types trace the
evolving large-scale clustering.

The data used here are formed by a subsample of the COSMOS2020
\cite{REF19} galaxy catalog. The filtering process was based on
photometric characteristics, selecting galaxies having $z\leq6$ and
magnitudes in the near-ultraviolet, optical, and near-infrared: NUV,
r, and K bands. The goal was to categorize galaxies into three color
types: blue, green and red. The results showed clear dependence of the
fractal dimension with galaxy colors:
$D_{\mathrm{blue}}~>~D_{\mathrm{red}}~>~D_{\mathrm{green}}$ having fractal
dimension varying in the range $D=1.40-2.03$ for $z<1$, and
$D_{\mathrm{blue}}~>~D_{\mathrm{green}}~>~D_{\mathrm{red}}$ in the
range $D=0.03-0.44$ for $1<z\leq4$. 

This paper is structured as follows. Sec.\ \ref{theory} presents the fractal
analysis methodology concerning the use of relativistic distances for the
attainment of the numerical densities of galaxies. Sec.\ \ref{data}
describes the COSMOS2020 survey and the data characteristics relevant to
this work. Sec.\ \ref{filter} describes the sample selection criteria,
including a tool for classifying galaxies by color using the fractal
dimension. Sec.\ \ref{resultados} presents the results and Sec.\
\ref{conc} is constituted by our conclusions.

\section{Fractal galaxy distribution}\lb{theory}

The theoretical basis of the fractal method capable of characterizing
how galaxies are irregularly distributed in space is conceptually very
simple and straightforward to apply. It is based on a fundamental
hypothesis for the empirical description of galaxy distribution called the
\textit{Pietronero-Wertz number distance relation}, which can be written as
follows,
\be
\Nobs=B \, {(\dobs)}^D. 
\lb{Nobs}
\ee
Here $\Nobs$ is the \textit{observed cumulative number counts} of galaxies,
$B$ is a positive constant, $\dobs$ is an \textit{observational distance},
and $D$ is the single fractal dimension. Eq.\ \eqref{Nobs} was first
advanced by Wertz \cite{wertz70,wertz71} under a different terminology.
Later it was independently rederived and interpreted under a fractal
framework by Pietronero \citetext{\citealp{pietronero87}; see also
\citealp{ribeiro98}, $\mathsection$III.4, and \citealp{ribeiro94},
$\mathsection$3}. As we shall show below this expression can be very
easily connected to other observational quantities
\citetext{see also \citealp{Teles2022}, and \citealp{Teles2023},
$\mathsection$1.2, app.\ B}.

Let $\Vobs$ be the \textit{observational volume} and $\gobs^\ast$ the
\textit{observed number density}, respectively written as follows,
\be
\Vobs=\frac{4}{3} \pi {(\dobs)}^3,
\lb{vobs}
\ee
\be
\gobs^\ast=\frac{\Nobs}{\Vobs}.
\lb{gobs-ast}
\ee
Now, substituting Eqs.\ (\ref{Nobs}) and (\ref{vobs}) into Eq.\
(\ref{gobs-ast}) we obtain the \textit{de Vaucouleurs density power-law}
\cite{vaucouleurs70}, 
\begin{equation}
\gobs^\ast = \frac{3B}{4\pi}{(\dobs)}^{D-3}.
\lb{gstar}
\end{equation}

The interplay among these quantities falls in two possible situations.
If $D~=~3$ both $\Nobs$ and $\Vobs$ grow at the same pace with increasing
values of $\dobs$, which then means that galaxies are evenly distributed
along the observed space as $\gobs^\ast$ remains constant. If $D~<~3$
then $\Nobs$ grows at a smaller rate than $\Vobs$, which then creates
gaps, or voids, in the galactic distribution as well as regions where
galaxies clump. This is so because the fractal dimension of these
structures depend on how $\Nobs$ grows cumulatively: its rate of
increase could vary with $\dobs$, which then affects the value of $D$
and cause $\gobs^\ast$ to vary. Hence, voids and galaxy clusters are a
by-product of the galaxy fractal system when $D$ is smaller than the
topological dimension where the galaxy structure is embedded. Eq.\
\eqref{gstar} provides the empirical method for detecting such a
variation between $\Nobs$ and $\dobs$ by measuring the power-law slope
in a log-log plot which in turn determines $D$.

Eqs.\ \eqref{Nobs} to \eqref{gstar} are straightforwardly applied to the
Newtonian cosmology approximation, but this is only possible up to
$z\approx0.2$ because beyond this range the various distance measures
differ for the same redshift value due to relativistic effects
\cite{ribeiro95,ribeiro2001b,ribeiro2005,juracy2008}. Previous works
dealing with observations under a fractal cosmology scenario have used
various cosmological distance definitions \cite{gabriela,Teles2022,
Teles2023}, but here it suffices to restrict ourselves only to the
\textit{luminosity distance} $\dl$ and \textit{galaxy area distance}
$\dg$, also known as \textit{transverse comoving distance}
\cite[$\mathsection$3.1]{ribeiro2001b}. They are connected by the
\textit{Etherington reciprocity law} below \citep{etherington33,
ellis2007},\footnote{Eq.\ \eqref{eth} is the second version of the
reciprocity law \cite{ellis2007}, whose first version contains the
\textit{observer area distance} (not used in this study). The first
version of the reciprocity law is also known in the literature as the 
\textit{distance duality relation} \cite{holanda2010}.}
\be
\dl=(1+z)\,\dg.
\lb{eth}
\ee
Therefore, the expressions above must be rewritten to become applicable for
$z \gtrsim 0.2$ as follows, 
\be
\dobs=d_i,
\lb{dists}
\ee
\be
\Vobs=V_i=\frac{4}{3} \pi {(d_i)}^3,
\lb{vi}
\ee
\be
\Nobs=N_i=B_i \, {(d_i)}^{D_i}, 
\lb{Nobs_i}
\ee
\be
\gobs^\ast=\gamma^\ast_i =\frac{N_i}{V_i}=\frac{3B_i}{4\pi}
{(d_i)}^{D_i-3},
\lb{gobs-ast_i}
\ee
where $i=({\ssty L},{\ssty G})$ according to the chosen distance
definition. Both the constant $B_i$ and the fractal dimension $D_i$
are now dependent on a specific distance measure because $N_i$ is
counted considering the limits given by each cosmological distance. So,
for a given $z$ each $d_i$ will produce its respective $V_i$, $N_i$,
$B_i$ and $D_i$.

Regarding the actual implementation of the above methodology with 
real data coming from galaxy redshift surveys, two remarks are noteworthy.
First, $\Nobs$ is viewed here as a radial average because this is how
fractal ideas are depicted in relativistic cosmological models
\cite{ribeiro92b,ribeiro2001b, bruno2025}. The second remark concerns
the existence of a decades old result coming from models of relativistic
fractal cosmologies which predicts that the number density, and so the
fractal dimension, will decrease for larger redshift values \cite[Fig.\
1]{ribeiro92b}, \cite[Figs.\ 1 and 3]{ribeiro95} and \cite[Fig.\ 2]
{ribeiro2001b}. In these models the average number density is defined
through radial averaging. Hence, the shift in $D$ to smaller values as
$z$ increases, particularly for $z>1$, is an effect to be expected
observationally. In fact, such prediction was already acknowledged in
previous works where the fractal dimension was derived through the
radial average of cumulative number counts from observational surveys
having deep galaxy redshift values \cite{gabriela, Teles2021,Teles2022}.
See also \cite[Figs.\ 3,4,9,10]{vinicius2007} and \cite[Figs.\ 7-12]
{iribarrem2012a}.

\section{COSMOS2020 Survey}\lb{data}

{The COSMOS2020 catalog \cite{REF19} is a multiwavelength dataset that
combines \textit{izYJHK} imaging from UltraVISTA Data Release 4 \cite{vizier},
$U$-band observations from the Canada-France-Hawaii-Telescope (CFHT)
MegaCam instrument \cite{REF20}, ultradeep optical data from Subaru's
Hyper Suprime-Cam (HSC) \cite{REF21}, and a comprehensive compilation
of mid-infrared imaging from Spitzer's Infrared Array Camera (IRAC)
obtained across the COSMOS field. The observed region covers
approximately 2 deg$^{2}$, spanning $1.54 \leq \mathrm{Dec},(\mathrm{deg})
\leq 2.88$ and $149.22 \leq \mathrm{RA},(\mathrm{deg}) \leq 150.81$.
Each new release of the catalog incorporates deeper observations and
more advanced data-processing techniques, improving the detection of
faint galaxies and extending the sample to higher redshifts. For
instance, the COSMOS2020 includes approximately 1.7 million galaxies,
reaching depths of $i \sim 27$, whereas COSMOS2015 \citep{Laigle2016}
contained about 500{,}000 galaxies down to $K_{s} \lesssim 24$.}

One of the key additions in COSMOS2020 is the secondary photometric
catalog known as FARMER, a profile-fitting extraction tool that
delivers self-consistent photometry without requiring PSF
homogenization or aperture corrections. This is an advantage when
combining images with different resolutions. As shown in Ref.\
\cite{REF19}, FARMER magnitudes are consistent with those obtained
using traditional SExtractor \citep{SExtractor}, but the former
provides more accurate photometry for fainter galaxies.

{As discussed in the previous section, our analysis relies on galaxy
number counts and cosmological distance estimates, making photometric
redshifts (photo-$z$) a key observational parameter. These redshifts
are derived through spectral energy distribution (SED) fitting,
obtained with LePHARE \cite{REF22} and EAZY \cite{REF23}, following
the methodology described in Ref.\ \cite{REF24}. In this work, we
adopted the LePHARE measurements derived for the FARMER photometry.}

\section{Data Selection}\lb{filter}

In order to minimize unreliable measurements we removed objects
with estimated stellar masses below $10^{7}$ M$_\odot$. Such very
low-mass systems are a challenge for characterization in galaxy
surveys and may arise from inappropriate assumptions in the adopted
star-formation histories or dust extinction models \citep{Mithi2025}.
In addition, we considered only galaxies with measurements in the
$NUV$, $r$ and $K$, as these are essential for the selection to be
described below. After applying these criteria our sample was reduced
to 852,988 galaxies.

\subsection{Magnitude limit}

Fractal studies of galaxy distributions depend on the use of
volume-limited samples. However, astronomical catalogs are typically
limited by apparent magnitude, which introduces a redshift-dependent
bias in the observed galaxy distribution. To mitigate this effect,
we adopted an observational efficiency threshold defined by an
absolute magnitude limit $M_{\mathrm{lim}}$ defined by
\begin{equation}
\label{eq6:eq6}
M_{\mathrm{lim}} = i - 5 \cdot \log(\dl)-25,
\end{equation}
where $i=27$ is the apparent magnitude in the $i$-band and $\dl$
is the luminosity distance in Mpc.

\begin{figure}[!ht]
    \centering
    \includegraphics[width=0.5\linewidth]{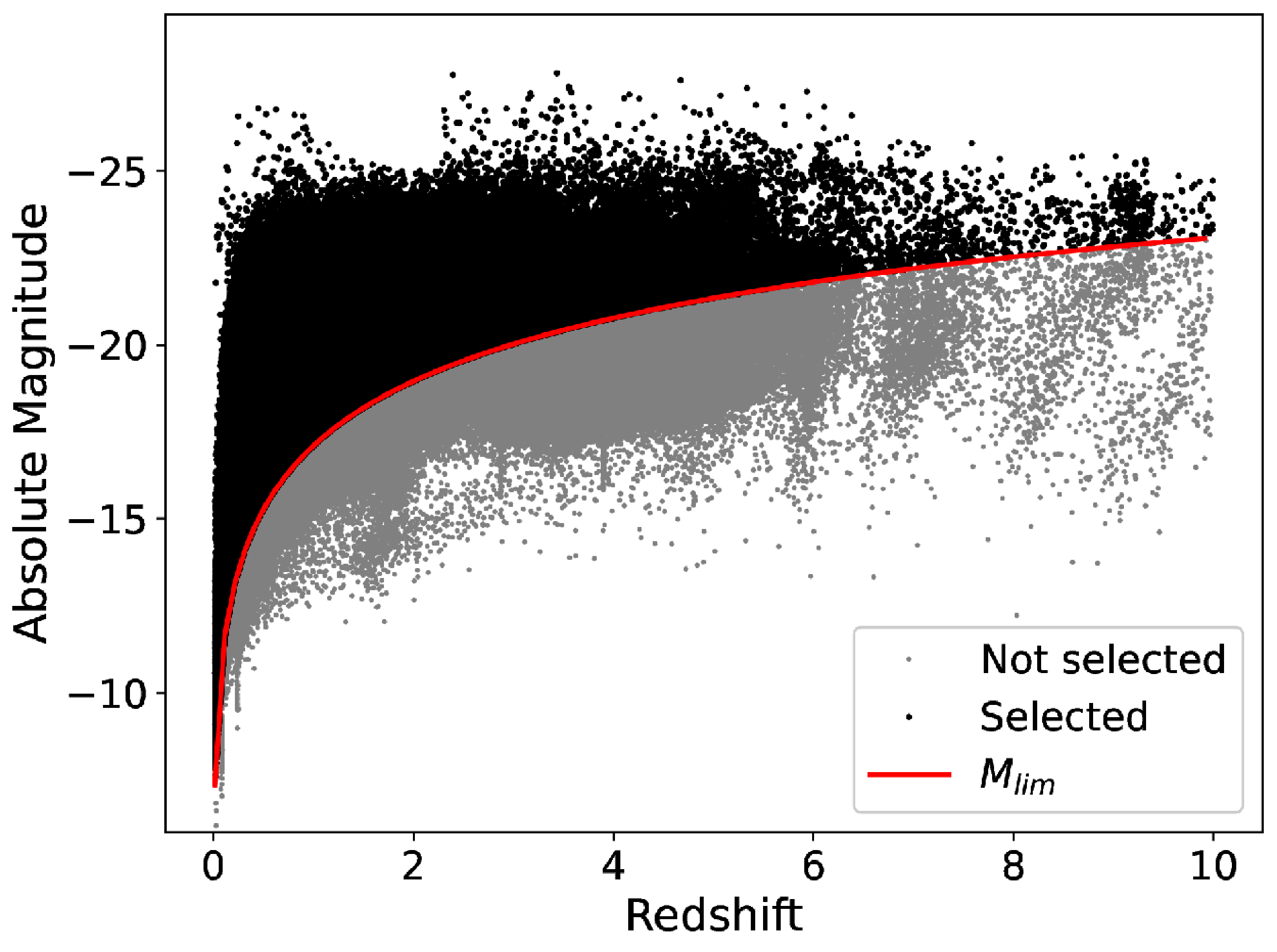}
    \caption{Absolute magnitude in the $i$-band as a function of
	     redshift for the initial sample of COSMOS2020. Black circles
	     indicate galaxies that lie above the red solid line, which
	     represents the limiting threshold $M_\mathrm{lim}$ defined
	     by Eq.\ \eqref{eq6:eq6}, while grey circles mark those that
	     fall below it and were discarded in this study.}
    \label{fig:fig1}
\end{figure}

Using absolute magnitudes in the $i$-band derived by the
LePHARE best-fit models, Fig.\ \ref{fig:fig1} shows the objects'
redshift evolution. Galaxies with absolute magnitudes brighter
than $M_{\mathrm{lim}}$ are marked in black and constituted
our reduced subsample. A small number of objects appear at the
extreme high-redshift tail of the distribution and, therefore,
we imposed an additional cut at $z=6$. Hence, the COSMOS2020
subsample was further reduced to 633,496 galaxies.

\subsection{Color-type populations}

\begin{figure}[bth]
    \centering
\includegraphics[width=1.0\linewidth]{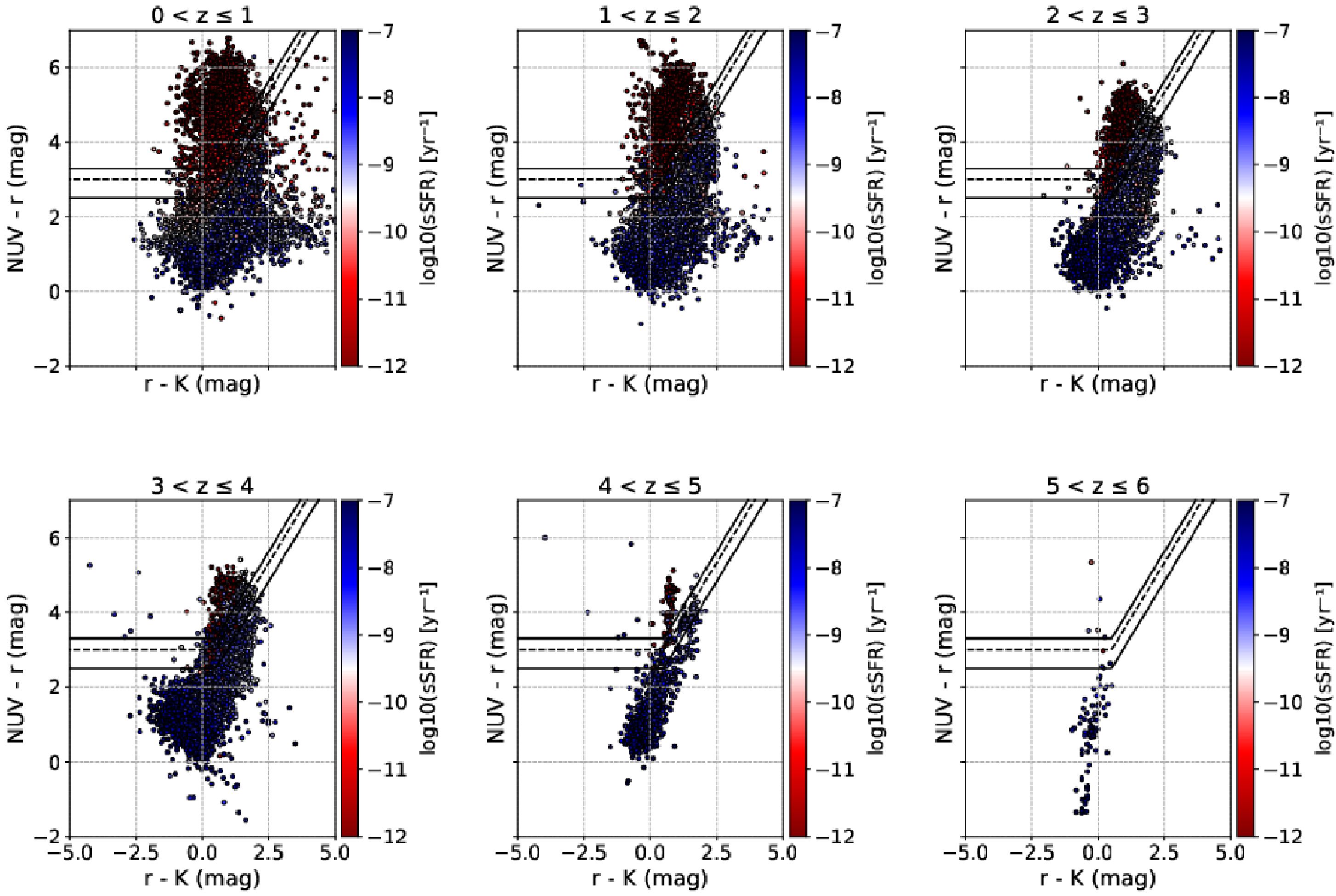}
\caption{$NUV-r$ vs. $r-K$ color-color diagrams for 6 equally lengthened
	redshift intervals. Solid lines correspond to the criterion defined
	by Ref.\ \cite[Eq.\ (2)]{REF26}, which separates galaxies into three
	classes: red (above the superior solid line), blue (below the
	inferior solid line), and green (in between the solid lines). The
        colorbar represents the sSFR.}
    \label{fig:fig2}
\end{figure}

{Color-color diagrams are one of the most traditional methodologies used
to separate different galaxy types such as the red sequence, blue cloud,
and green valley \citep[e.g.][]{Daddi2004,Arnouts2013,Moutard2016,Vergani2018}.
Blue galaxies are typically star-forming systems with relatively high specific
star formation rates (sSFR; $\log \mathrm{sSFR} > -9$), red galaxies are
quiescent with low sSFR ($\log \mathrm{sSFR} < -10$), and green valley
galaxies represent transitional objects with intermediate properties.
Following the classifications defined by Ref.\ \cite[Eq.\ (2)]{REF26},
we used absolute magnitudes in the $NUV$, $r$, and $K$ bands to construct
$(NUV - r)$ versus $(r - K)$ color--color diagrams and select each sample
of galaxy type. In total, the sample includes 544,814 blue galaxies and
51,226 red galaxies, while 37,456 objects fall within the green-valley
population. Figure~\ref{fig:fig2} shows these diagrams, highlighting the
three galaxy categories across six redshift intervals: $(z \leq 1)$,
$(1 < z \leq 2)$, $(2 < z \leq 3)$, $(3 < z \leq 4)$, $(4 < z \leq 5)$,
and $(5 < z \leq 6)$.} 

\begin{figure}[bth]
    \centering
    \includegraphics[width=0.49\linewidth]{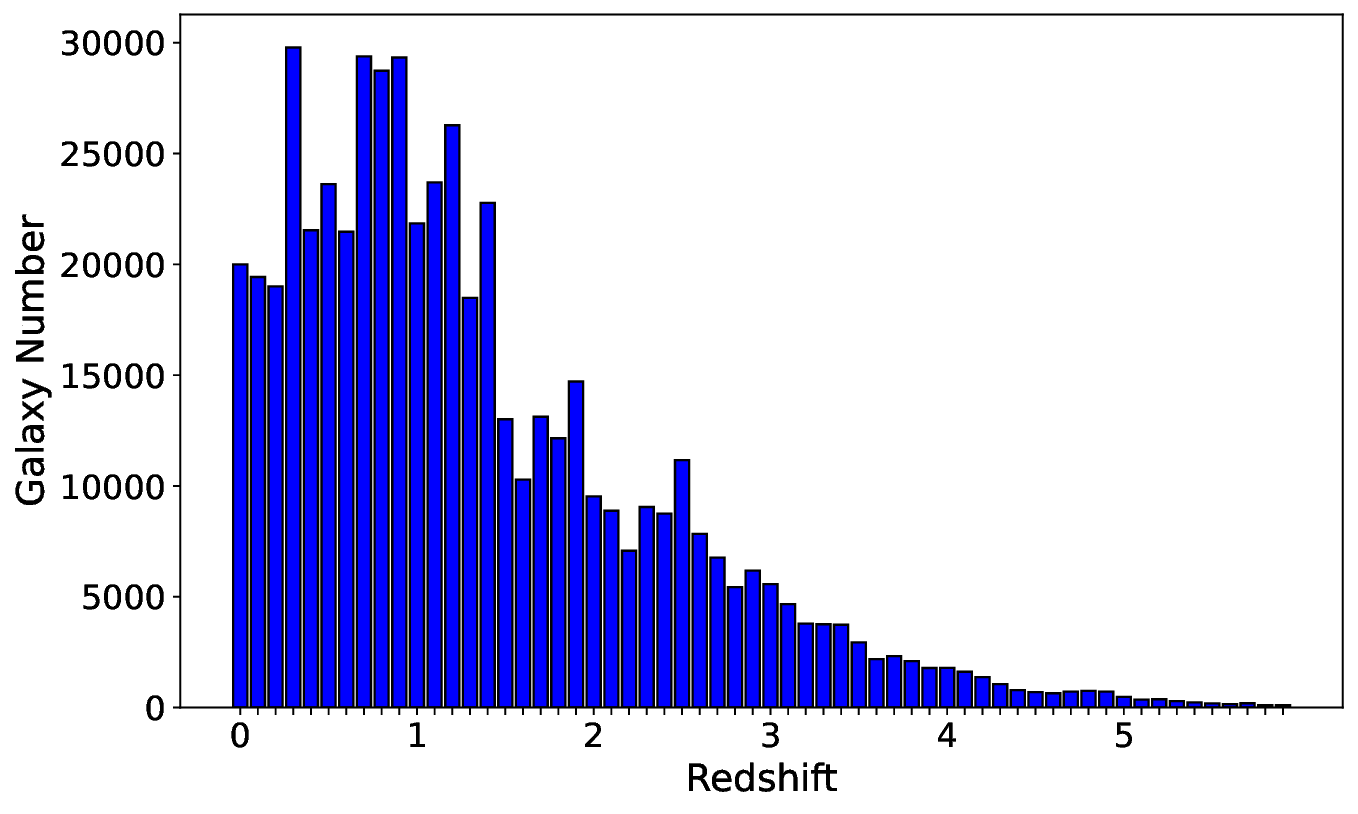}
    \includegraphics[width=0.49\linewidth]{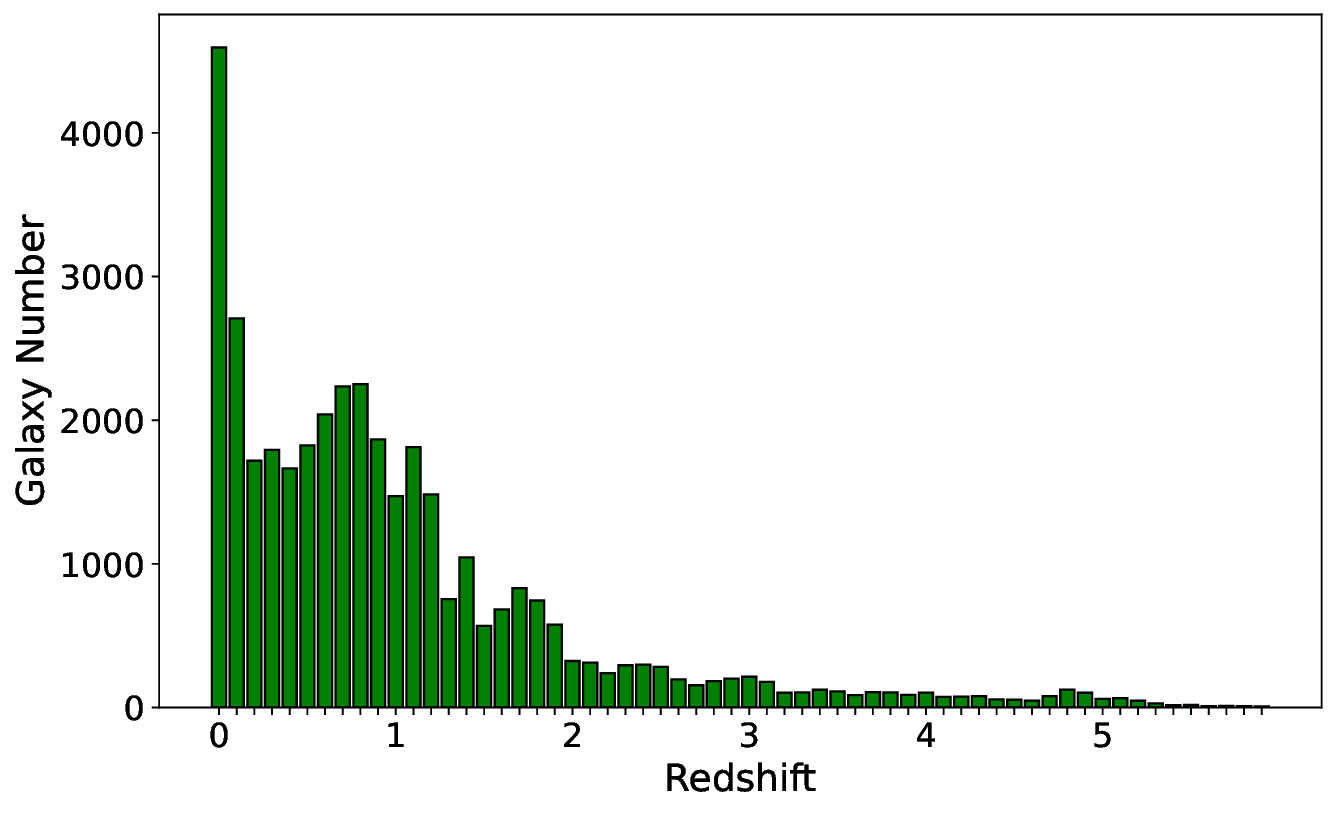}
    \includegraphics[width=0.49\linewidth]{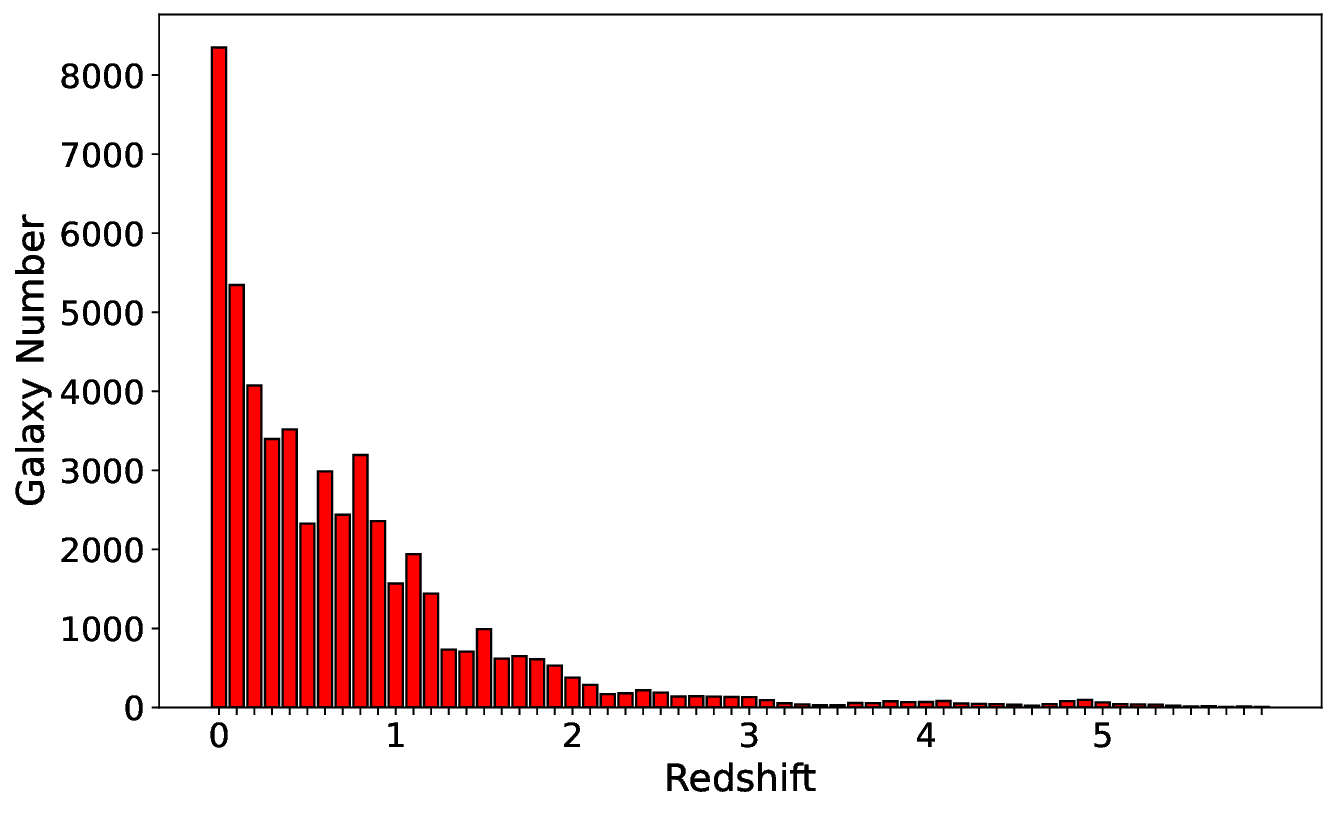}
    \caption{Histogram of the redshift distribution of blue (top left),
	    green valley (top right) and red galaxy types (bottom). After
	    discarding galaxies having $z>4$ we ended up with a final
	    subsample containing 618,952 objects, constituted by 532,190
	    blue galaxies, 50,383 red ones, and 36,379 objects being
	    classified as galaxies belonging to the green-valley.}
    \label{fig:fig3}
\end{figure}

{By examining Fig.\ \ref{fig:fig2} and the galaxy redshift distribution
in Fig.\ \ref{fig:fig3}, one can notice that for $z > 4$ the number of
detected objects drops sharply across all galaxy types. For this reason
we further restricted our analysis and discussion to the range $z\leq4$,
leading to a final subsample containing 618,952 galaxies.}

\section{Results}\lb{resultados}

\begin{figure}[ht]
    \centering
    \includegraphics[width=0.49\linewidth]{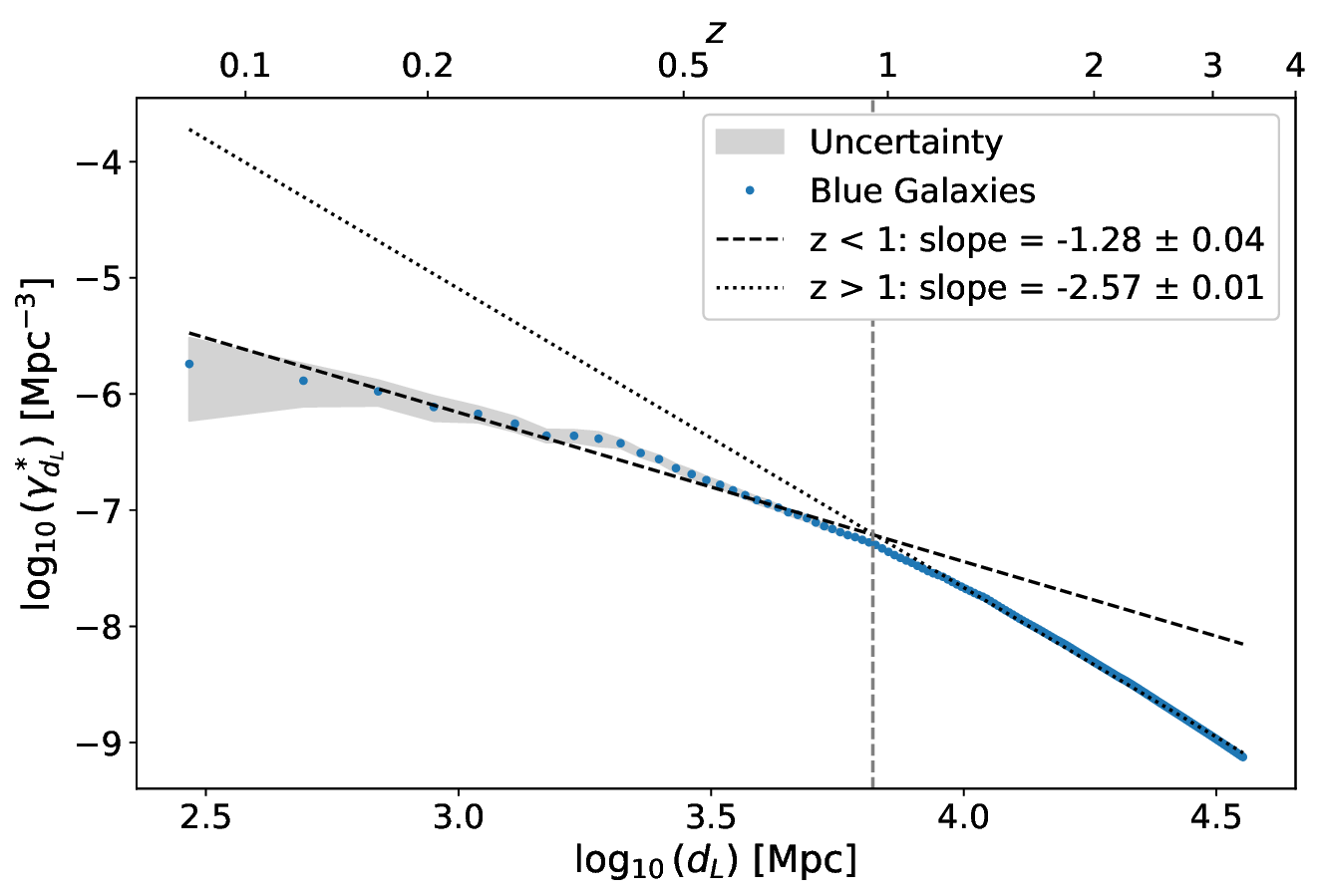}
    \includegraphics[width=0.49\linewidth]{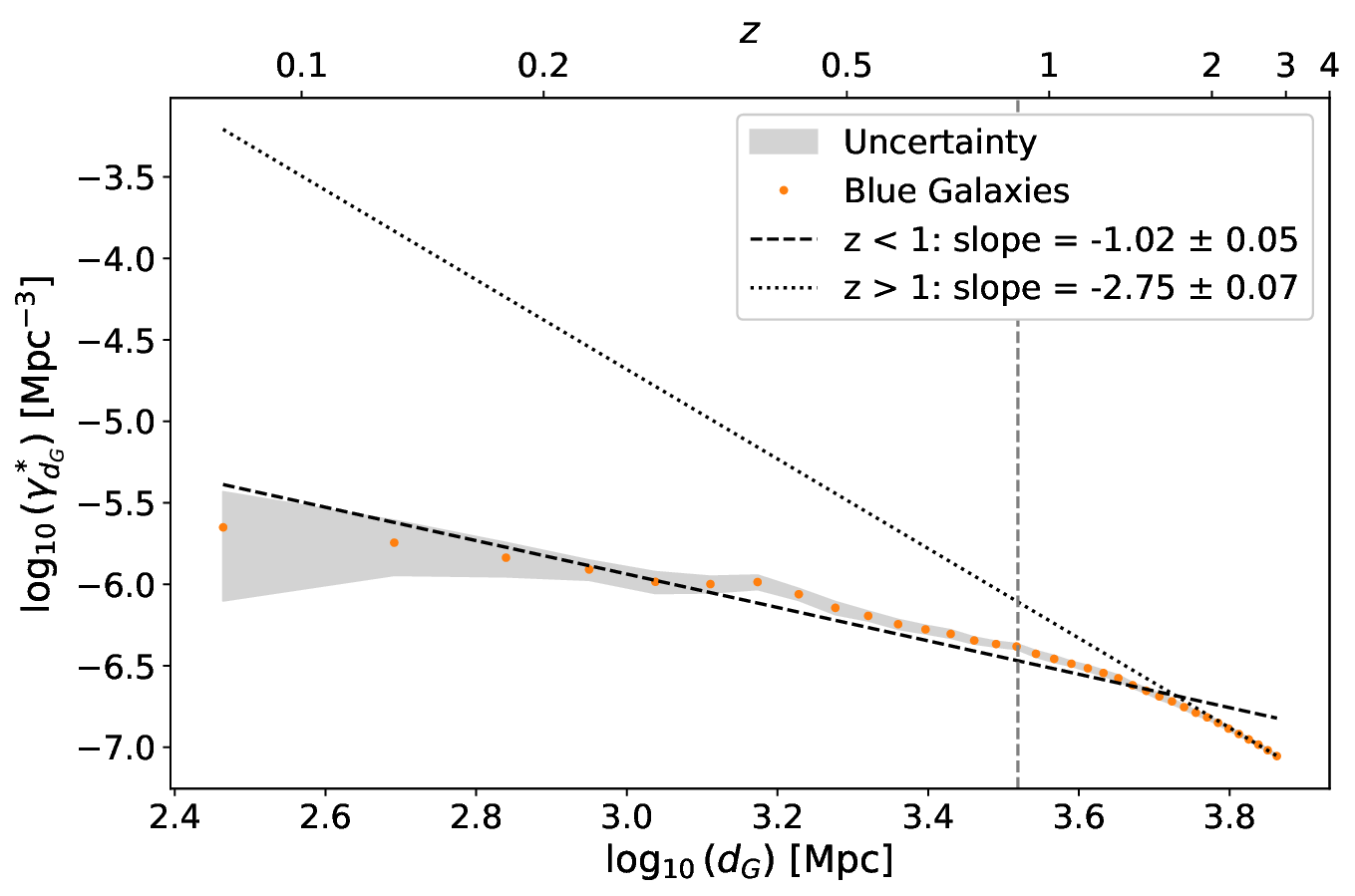}
    \includegraphics[width=0.49\linewidth]{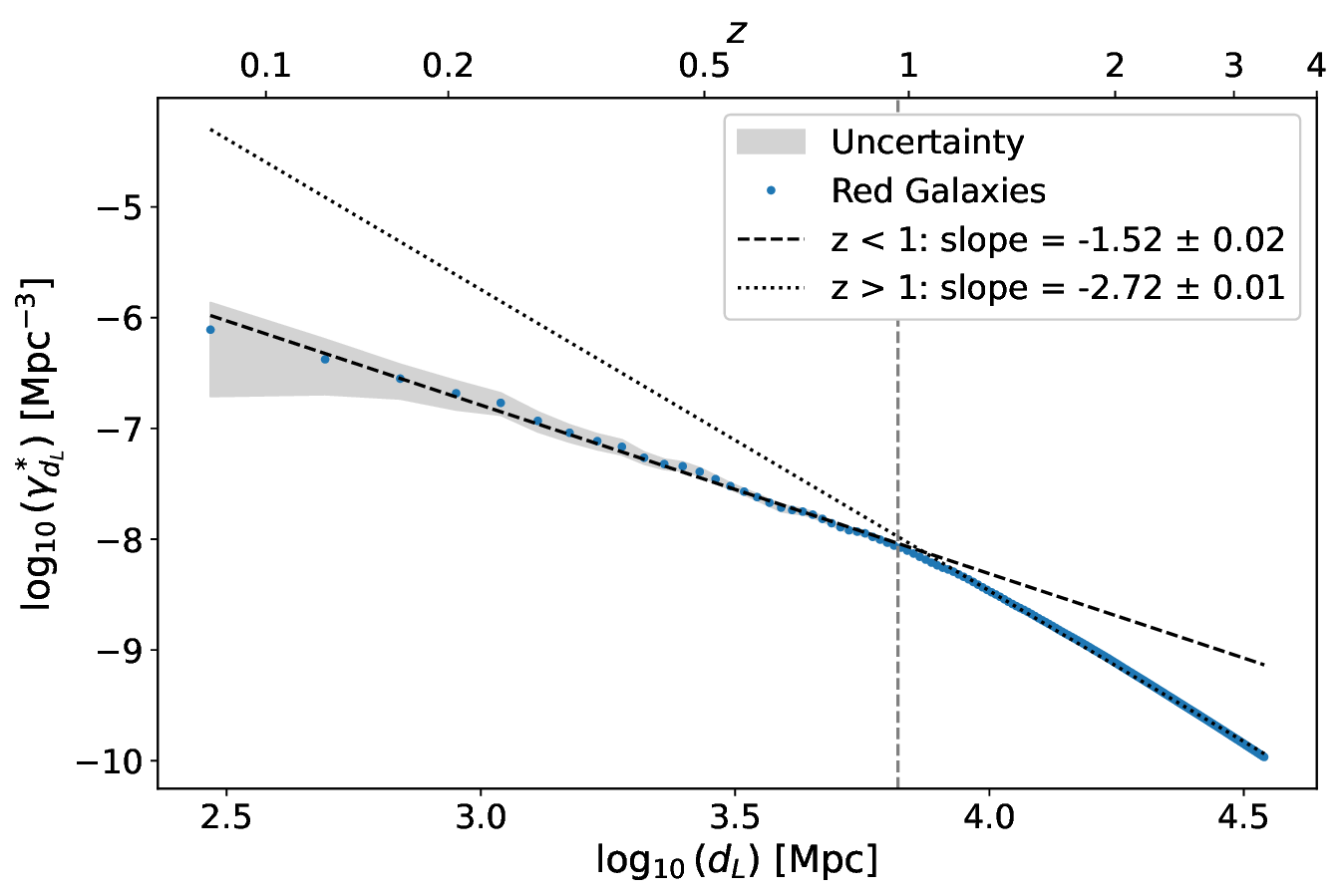}
    \includegraphics[width=0.49\linewidth]{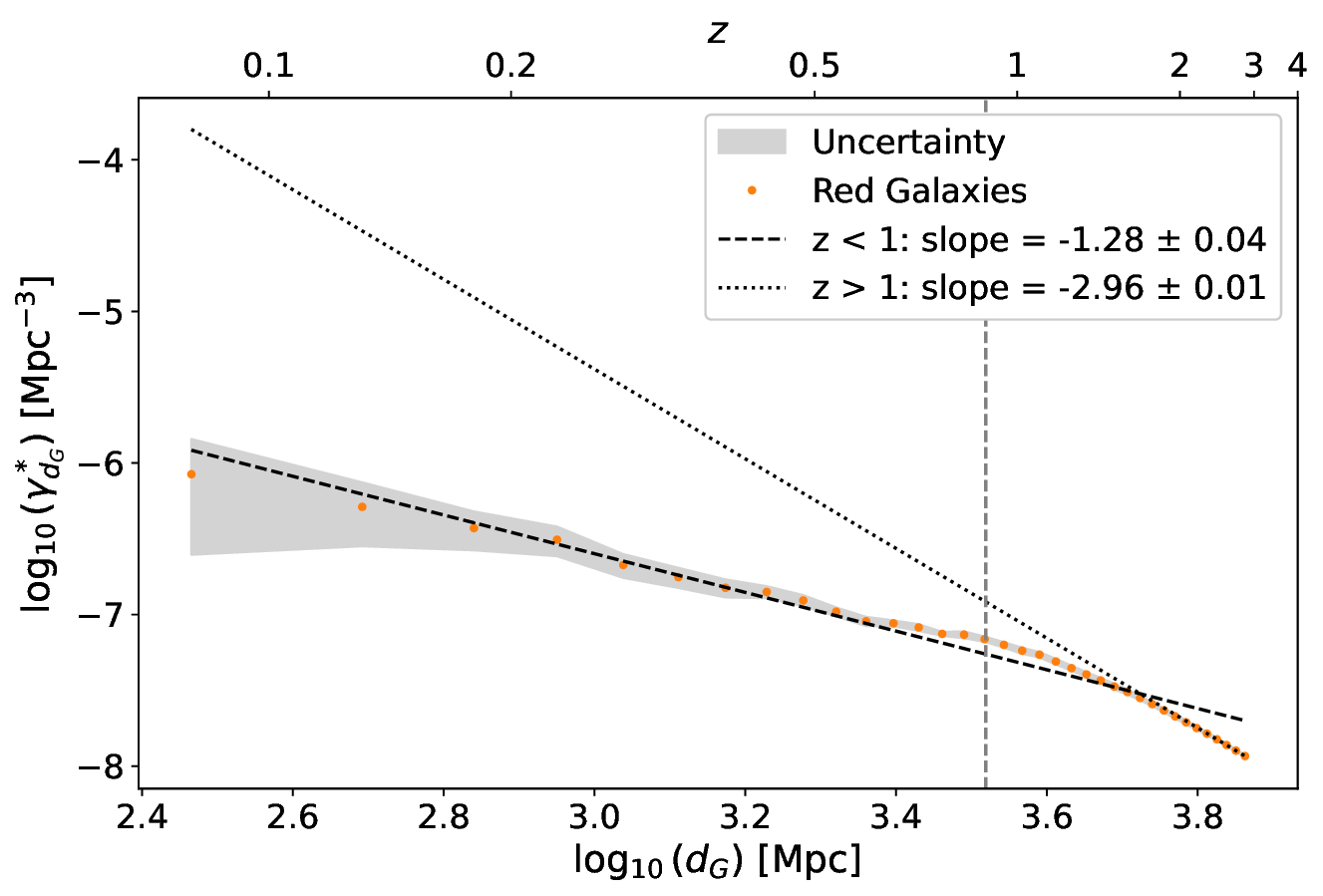}
    \includegraphics[width=0.49\linewidth]{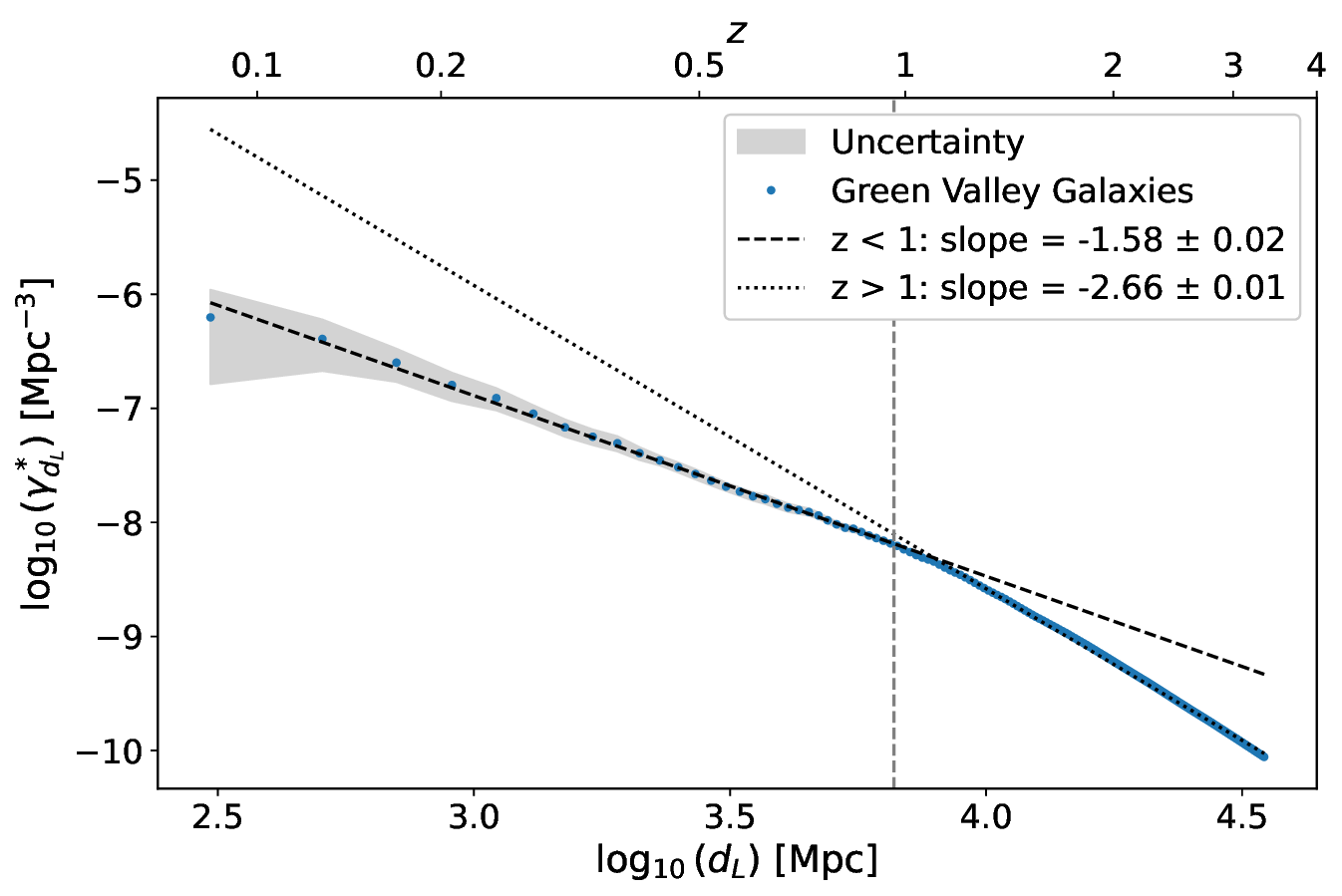}
    \includegraphics[width=0.49\linewidth]{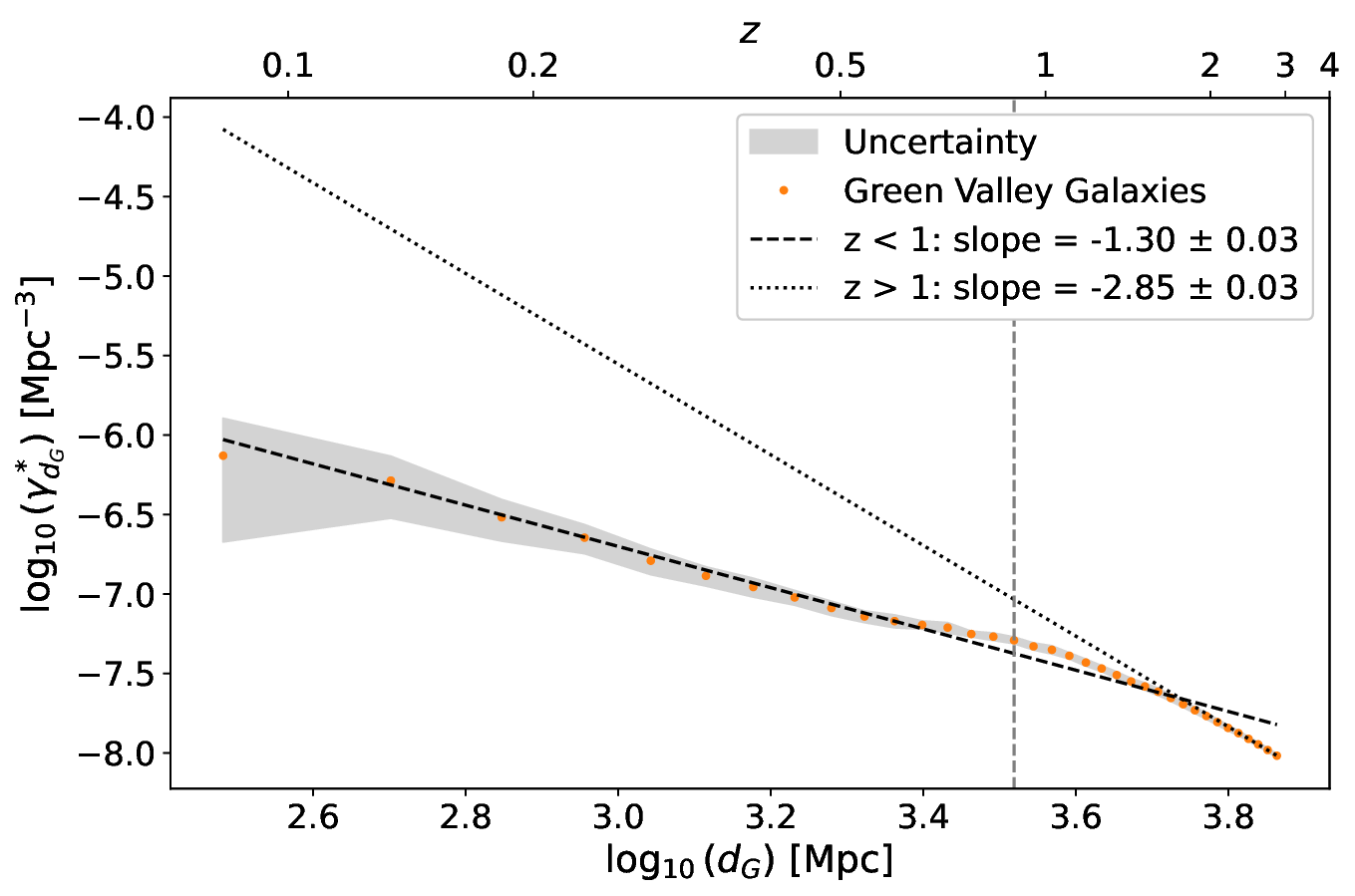}
    \caption{Galaxy number densities $\gobs^\ast$ obtained as counts per
    cumulative
    volume with steps of 200 Mpc in terms of the luminosity distance $\dl$
    and galaxy area distance $\dg$ \cite[$\mathsection$2]{ribeiro2005}
    for three color type populations:  blue, red and green. The shaded
    gray region represents the 1$\sigma$ uncertainty range derived from
    the upper and lower bounds of the photometric redshifts (lpzPDFu68,
    lpzPDFl68). The vertical dashed line marks the fractal dimension
    transition at $z = 1$.}
    \label{fig:enter-label}
\end{figure}

Analysis of the observed number density $\gamma^\ast$ in terms of the
distances measures $\dl$ and $\dg$ in log-log plots shown in Fig.\
\ref{fig:enter-label} revealed two well-defined power-law intervals
separated by the observational boundary at $z=1$. The slopes obtained
through linear regression were converted into fractal dimensions using
the relation $D=3+\mathrm{slope}$, resulting in different values for
$D$ according to the galaxy color type.

For \textit{blue} galaxies we obtained $\Dl=1.72\pm0.04$ and $\Dg=1.98
\pm0.05$ at $z<1$, and $\Dl=0.43\pm0.01$ and $\Dg=0.25\pm0.07$ in the
range $1<z\leq4$. \textit{Red} galaxies exhibited lower initial dimensions,
having $\Dl=1.48\pm0.02$ and $\Dg=1.72 \pm 0.04$ for $z<1$, which drop
to $\Dl=0.28\pm0.01$ and $\Dg=0.04\pm0.01$ for $1<z\leq4$. Finally, the
\textit{green valley} galaxies yielded $\Dl = 1.42\pm0.02$ and $\Dg=1.70
\pm0.03$ for $z<1$, decreasing to $\Dl=0.34\pm0.01$ and $\Dg=0.15\pm0.03$
at the higher redshift interval of $1<z\leq4$. These results are collected
in Tab.\ \ref{tab1}.
\begin{table}[ht]
\centering
\begin{tabular}{l
                S[table-format=1.2(2)]
                S[table-format=1.2(2)]
                S[table-format=1.2(2)]
                S[table-format=1.2(2)]}
\toprule
\textbf{Color} 
& {$\Dl \; (z<1)$} 
& {$\Dl \; (1<z\leq4)$} 
& {$\Dg \; (z<1)$} 
& {$\Dg \; (1<z\leq4)$} \\
\midrule
Blue  & 1.72 \pm 0.04 & 0.43 \pm 0.01 & 1.98 \pm 0.05 & 0.25 \pm 0.07 \\
Red   & 1.48 \pm 0.02 & 0.28 \pm 0.01 & 1.72 \pm 0.04 & 0.04 \pm 0.01 \\
Green & 1.42 \pm 0.02 & 0.34 \pm 0.01 & 1.70 \pm 0.03 & 0.15 \pm 0.03 \\
\bottomrule
\end{tabular}
\caption{Results in two redshift scales of the COSMOS2020 galaxy survey
fractal analysis in the reduced subsample shown by color. The single
fractal dimensions $\Dl$ and $\Dg$ were obtained through the luminosity
distance $\dl$ and galaxy area distance $\dg$, also known as transverse
comoving distance, respectively. The very low values $\Dl=0.28\pm0.01$
and $\Dg=0.04\pm0.01$ for red galaxies at the higher redshift interval
of $1<z\leq4$, the lowest fractal dimension results found in this study,
indicate the virtual absence of these galaxy color types in that redshift
range, as shown in Fig.\ \ref{fig:fig3}}
\lb{tab1}
\end{table}

The consistency of this color type distinction even after accounting for
photometric redshift error propagation suggests that the fractal dimension
does provide an observational parameter capable of distinguishing different
galaxy population types in terms of color. This interpretation is
reinforced by the fact that the blue population, dominated by recent star
formation, consistently exhibits the highest fractal dimensions, indicating
a relatively dense spatial filling. In contrast, quiescent galaxies occupy
regions of lower global density, resulting in lower dimensional values. The
intermediate position of green valley galaxies suggests that the decline in
star formation rate is already affecting the large-scale topology, albeit
less markedly than in red galaxies.

This pattern replicates with greater resolution and detail the blue-red
contrast observed in previous surveys, namely COSMOS2015, SPLASH, and
UltraVISTA \cite{Teles2021,Teles2022}, which showed that each
population displays a characteristic fractal dimensional ``signature.''
The results obtained here (see Tab.\ \ref{tab1}) present a clear gradient:
$D_{\mathrm{blue}}~>~D_{\mathrm{red}}~>~D_{\mathrm{green}}$ for $z<1$, and
$D_{\mathrm{blue}}~>~D_{\mathrm{green}}~>~D_{\mathrm{red}}$ for $1<z\leq4$.
Such ordering for blue and red color galaxy populations was previously
pointed out by Ref.\ \cite{Teles2022} that the fractal dimension can be
used as a descriptor of an intrinsic property, remaining stable for a
star-forming populations but contracting significantly in quiescent ones.
Thus, the present dataset reinforces the central proposition of this
work, namely that $D$ seems to provide a descriptive indicator of color
differentiation which might mean distinguishing evolutionary galaxy classes.

In addition to internal consistency, our values align with the fractal
dimension ranges of $1.42-1.83$ for $z<1$ and $0.23-0.81$ for $1<z\leq6$
obtained with the UltraVISTA DR1 using similar methodologies
\cite{Teles2021}, but without galaxy distinction by color. The recurrence
of a ``dimensional spectrum'' associated with galaxy colors or star
formation rates suggests that $D$ captures physical information that
complements traditional photometric indices. In particular, it quantifies
the degree of spatial clustering through a single quantity that appears to
remain stable under moderate changes in magnitude cuts and in the choice
of cosmological distance measures.

Some limitations should be noted. The dimensions for $z > 1$ are affected
by sparse statistics, especially within the red galaxy samples, and
by the propagation of photometric redshift uncertainties. Deeper
spectroscopic surveys would be required to validate the persistence of
these differences at earlier cosmic times. Additionally, the employed
method relies on a single fractal dimension approach. Therefore,
multifractal extensions could reveal internal variations in $D$ within
the same category and further enhance the method as a classification tool.

\section{Conclusions}\lb{conc}

This work investigated the applicability of the single fractal dimension $D$
as a parameter for distinguishing galaxy populations by color using 
large-scale redshift survey data. The subsample of 618,952 galaxies from
the COSMOS2020 survey catalog, whose selection was based on strict photometric
and completeness criteria in the $NUV$, $r$, and $K$ bands up to $z=4$,
enabled the identification of three distinct galaxy populations by color
types, namely, blue, green, and red. This was done through color-color
diagrams depicting star formation rates.

Cumulative number counts were assumed to be a function of cosmological
distances by means of the \textit{Pietronero-Wertz number distance
relation}, which in turn allowed us to write the \textit{de Vaucouleurs
density power-law} that relates galaxy number densities to distances.
The single fractal dimension $D$ were then derived from plots constituted
by the linearized slope of the power-law.

Fractal dimensions obtained this way from the above mentioned COSMOS2020
survey subsample revealed systematic variations correlated with galaxy
color types. Star-forming, or blue color galaxies, consistently presented
the highest fractal dimensions, indicating a denser spatial distribution,
while low star-formation, or red color galaxies, showed lower dimensions,
suggesting that these galaxies are found in less dense regions. Green
color galaxies showed approximate intermediary values for higher redshift
values, possibly suggesting them as being a transition between active and
inactive types.

The results showed two fractal dimensional gradients, namely,
$D_{\mathrm{blue}}~>~D_{\mathrm{red}}~>~D_{\mathrm{green}}$ for $z<1$, and
$D_{\mathrm{blue}}~>~D_{\mathrm{green}}~>~D_{\mathrm{red}}$ for $1<z\leq4$,
gradient which showed to be robust even when accounting for photometric
redshift uncertainties. These results reinforce previous findings by
Ref.\ \cite{Teles2022} who showed fractal dimension dependency on 
galaxy color types. Despite data limitations at high redshifts,
especially for the red population, and the use of single slope models,
our analysis supports the use of fractal dimension as a parameter to
discriminate galaxies by their colors.

In summary, the fractal geometry methodology presented here seems to
offer a straightforward quantitative way to track observational aspects
of large-scale galaxy populations that goes beyond the discussion if
the large-scale structure of the Universe eventually turns into a
homogeneous distribution, a subject that has so far constrained the
application of fractal geometry in cosmological studies. Hence, fractal
methods such as the one exposed here could in principle be applied to
other observational galaxy features, because they provide a sensitive
diagnostic for how galaxies trace the evolving cosmic web.

\section*{Acknowledgments}

S.T.\ acknowledges financial support from \textit{Coordena{\c{c}\~{a}}o
de Aperfei{\c{c}}oamento de Pessoal de N\'{\i}vel Superior - Brasil} (CAPES)
- Finance Code 001.  A.R.L acknowledges the grant number 2025/09544-0 from 
\textit{S\~{a}o Paulo Research Foundation} (FAPESP) and financial support 
from \textit{Consejo Nacional de Investigaciones Cientificas y T\'{e}cnicas}
(CONICET). M.B.R.\ received partial financial support from the \textit{Rio de 
Janeiro State Research Funding Agency} (FAPERJ), grant number
E-26/210.552/2024.

\section*{Author contribution}
A.E.L., J.C.S., A.C.S.T.\ and M.V.T.\ developed the Python code for data
selection, reduction and analysis, and produced the first text draft.
S.T.\ and A.R.L.\ supervised and contributed to the code development.
A.R.L.\ and M.B.R.\ provided the theoretical underpinning, supervised
the whole research and edited the final text.

\section*{Data Availability Statement}
This manuscript has no associated data.

\section*{Code Availability Statement}
This manuscript has no associated code/software. 

\section*{Conflict of interest}
The authors declare that they have no known competing financial interests
or personal relationships that could have appeared to influence the work
reported in this paper. 













\bibliography{cosmo}
\bibliographystyle{elsarticle-num}
\end{document}